\documentclass[a4paper,11pt]{article}
\usepackage{pos}
\usepackage{multirow}
\usepackage{makecell}
\usepackage{microtype}
\usepackage{upgreek}

\renewcommand{\Re}{{\rm Re}\,}
\renewcommand{\Im}{{\rm Im}\,}
\renewcommand{\bar}[1]{\mkern 1.5mu\overline{\mkern-1.5mu#1\mkern-1.5mu}\mkern 1.5mu}

\title{QCD Anderson transition with overlap valence quarks on a twisted-mass sea -- an update}
\author*[a]{Robin Kehr}
\author[a,b]{Lorenz von Smekal}

\affiliation[a]{Institut f\"ur Theoretische Physik, Justus-Liebig-Universit\"at,\\
Heinrich-Buff-Ring 16, 35392 Giessen, Germany}

\affiliation[b]{Helmholtz Forschungsakademie Hessen f\"ur FAIR (HFHF), GSI Helmholtzzentrum f\"ur Schwerionenforschung, Campus Gießen}

\emailAdd{Robin.Kehr@physik.uni-giessen.de}
\emailAdd{lorenz.smekal@physik.uni-giessen.de}

\abstract{We investigate the QCD Anderson transition by studying the low-lying eigenmodes of the overlap operator in the background of gauge configurations with 2+1+1 quark flavors of twisted-mass Wilson fermions. The mobility edge, below which eigenmodes are localized, is estimated by the inflection point of the relative volume. The analysis of its temperature dependence suggests a close relation of localization to chiral symmetry restoration. We update our previous work \cite{Kehr:2023wrs} by including recent results on lower temperatures and switching to improved estimates of the lattice spacing and pseudocritical temperature respectively pion mass. Contrary to the previous prediction, our the mobility edge estimate does not vanish at the temperature of the chiral phase transition. We discuss a possible scenario, supported by literature, for why this could be the case.}

\FullConference{The 41st International Symposium on Lattice Field Theory (LATTICE2024)\\
 28 July - 3 August 2024\\
Liverpool, UK\\}

\begin{document}

\maketitle

\section{Introduction}

The term Anderson transition originates from condensed matter physics, where it describes the metal-insulator transition in disordered solids \cite{Anderson:1958vr, Evers:2007zsx}. In the metal phase the low-lying eigenmodes of the Hamiltonian are delocalized and give rise to conductivity, whereas above some critical disorder strength all eigenmodes are localized making the conductivity vanish. The energy threshold, which seperates localized and delocalized modes, is called mobility edge. In QCD, there is an analogous transition, where however the low-lying eigenmodes of the Dirac operator are localized and the higher are delocalized above some temperature. Below this temperature, all modes are delocalized \cite{Giordano:2021qav}. Thus, the temperature takes over the role of the disorder strength, which seems be related to the Polyakov loop, since it was found that the eigenmodes tend to localize in the sinks of the Polyakov loop \cite{Holicki:2018sms}. As a consequence, the vanishing of the mobility edge coincides with the deconfining phase transition in quenched QCD \cite{Kovacs:2017uiz}. In QCD setups with dynamical fermions, it was also found that the same occurs at the chiral transition \cite{Giordano:2014pfa, Holicki:2018sms} and at the chiral phase transition temperature in the chiral limit in our previous work \cite{Kehr:2023wrs}. Additionally, in Ref.~\cite{Giordano:2022ghy}, it was argued that in the chiral limit no goldstone bosons exist if near-zero modes are localized. On top of that, the near-zero are directly connected to chiral symmetry breaking as the produce the chiral condensate via the Banks-Casher relation \cite{Banks:1979yr}. Summarizing all of this, studying the QCD Anderson transition might provide the answer to the question whether and how chiral and deconfinement transition are related to each other.

This work includes recent results on lower temperatures of the most physical data set evaluated in Ref.~\cite{Kehr:2023wrs}. We compute the low-lying eigenmodes of the overlap operator
\begin{equation}
	D_{\mathrm{ov}} = \frac{\rho}{a} \, (1 + \mathrm{sgn}\,K) \,,
\end{equation}
which was introduced in Ref.~\cite{Neuberger:1997fp} and implements chiral symmetry on the lattice by satisfying the Ginsparg-Wilson relation proposed in Ref.~\cite{Ginsparg:1981bj}. $K$ represents the Wilson operator with a negative mass term $-\rho$, where parameter $\rho$ was to $1.4$ in order to optimize locality, which was found to be the optimal choice according to Ref.~\cite{Cichy:2012vg}. The overlap operator was implemented using a rational approximation and the eigenmodes were computed with the \emph{Krylov-Schur method} of the SLEPc library \cite{slepc}. As background we employ gauge configurations of the \emph{twisted mass at finite temperature} (tmfT) collaboration, which were generated with $N_\mathrm{f} = 2+1+1$ flavors (two degenerate light, physical strange \& charm quarks) twisted mass Wilson fermions at maximal twist and the Iwasaki gauge action. Table~\ref{tab:overview_eigenmodes} provides an overview of the computed eigenmodes with the data to $N_\mathrm{t} \in \{14,16,18,20,24\}$ being the recently produced results. Compared to \cite{Kehr:2023wrs}, the lattice spacing $a = 0.0619(18)\,\mathrm{fm}$ and the pseudocritical temperature $T_\mathrm{pc}=171(6)\,\mathrm{MeV}$ of the chiral transition respectively pion mass $m_\uppi = 225(7)\,\mathrm{MeV}$ were updated according to the more recent determinations in Ref.~\cite{ExtendedTwistedMass:2019omo} and \cite{Kotov:2021rah}, which were taken from \cite{Alexandrou:2014sha} and \cite{Burger:2018fvb} in the past. As a result, all three parameters became slightly larger and especially the estimate for $T_\mathrm{pc}$ is improved due to the determination by an asymmetric fit to the chiral susceptibility. The number of lattice sites in each space direction amounts $N_\mathrm{s}=48$, such that the lattice extent is given by $L = 2.97(9)\,\mathrm{fm}$.

As a means of estimating the effect of continuum extrapolations, we apply the stereographic projection
\begin{equation}
	\lambda^\prime:= \frac{\mathrm{i}\, \Im \lambda}{1 - \frac{a}{2\rho} \Re \lambda} 
\end{equation}
to the eigenvalues $\lambda$, which are distributed on the Ginsparg-Wilson circle. We define the relative volume an eigenmode occupies
\begin{equation}
	r(\lambda) = \frac{P_2^{-1}(\lambda)}{|\Lambda|} \in [1/|\Lambda|,1] \label{eq:eigenmode_relvol}
\end{equation}
as measure of localization, where the $q$-th order inverse participation ratio of an eigenmode $v_\lambda$ to the eigenvalue $\lambda$ reads
\begin{equation}
	P_q(\lambda)  = \sum_{i \in \Lambda} ( v_\lambda(i)^\dagger  v_\lambda(i))^q 
\end{equation}
and $|\Lambda|$ denotes the total number of lattice sites.

As a proxy for the mobility edge we employ the inflection point $\lambda_\mathrm{c}$ of the relative point volume, which we extract as follows: In order to approximate a smooth function we average $\lambda$ and $r(\lambda)$ over small bins of size $\Delta\lambda$. For better readability we therefore redefine  $\lambda:=\bar{|\lambda^\prime|}$ and $r(\lambda):=\bar{r(\lambda)}$ in the following. We then fit the Taylor polynomial 
\begin{equation}
	r(\lambda) = r_\mathrm{c} + b \, (\lambda - \lambda_\mathrm{c}) + c \, (\lambda - \lambda_\mathrm{c})^3 + d \, (\lambda - \lambda_\mathrm{c})^4
\end{equation}
to the data and vary the fit interval $[\lambda_{\mathrm l}, \lambda_{\mathrm r}]$ and the binsize to obtain fits with $\chi^2 / \mathrm{d.o.f.} \approx 1$. The results for $\lambda_\mathrm{c}$, which is the main quantity of interest, are listed in Table~\ref{tab:overview_eigenmodes} as well. As improvement to \cite{Kehr:2023wrs}, additionally the systematic error of $\lambda_\mathrm{c}$ was estimated by the deviation to the inflection point of the second best fit. The same analysis can be done with respect to the Ginsparg-Wilson angle instead of the stereographic projection but it was found out in the past that this just gives rise to minor differences of the results. Since it is believed that the stereographic projection brings us closer to the continuum limit, we only analyzed this variant here.  

\begin{table}
	\centering
	\begin{tabular}{|c|c|c|c|c|c|c|}
		\hline
		\Gape[7pt][7pt] Set of ensembles & $N_\mathrm{t}$ & $T$ / $\mathrm{MeV}$ & $T/T_\mathrm{pc}$ & $\lambda_\mathrm{c}$ / $\mathrm{MeV}$ & \# $\mathrm{conf.}$ & $\frac{\mathrm{modes}}{\mathrm{conf.}}$	\\
		\hline\rule{0pt}{0.86\normalbaselineskip}
		\multirow{10}{*}{\makecell{\textbf{D210} \\ $N_\mathrm{s}=48$ \\ $a = 0.0619(18)\,\mathrm{fm}$ \\ $m_\uppi=225(7)\,\mathrm{MeV}$ \\ $T_\mathrm{pc}=171(6)\,\mathrm{MeV}$}}
		&	4	&	797(23)	&	4.66(23)	&	3091(90)(4)(2)	&	10	&	1000	\\
		&   6   &	531(15) &   3.11(16)	&	1638(48)(9)(6)	&   10  &   700		\\
		&   8   &	398(12) &   2.33(12)    &	971(28)(4)(1)	&   10  &   500		\\
		&   10 	&	319(9)  &   1.86(9)     &   644(19)(8)(16)  &   10  &   400		\\
		&	12 	&	266(8)  &   1.55(8) 	&   452(13)(22)(11) &   10  &   350		\\
		&   14  &	228(7)  &   1.33(7)     &	255(7)(5)(7)	&   10  &   300		\\
		&   16  &	199(6)  &   1.17(6)     &	151(4)(10)(2)	&   10  &   250		\\
		&   18  &	177(5)  &   1.04(5)     &	107(3)(8)(22)	&   10  &   225		\\
		&   20  &	159(5)  &   0.93(5)     &	86(3)(3)(5)     &   10  &   200		\\
		&   24	&	133(4)  &   0.78(4)     &	88(3)(5)(18)	&   6  	&   175		\\
		\hline
	\end{tabular}
	\caption{List of tmfT ensembles with numbers of configurations on which overlap eigenmodes were computed and numbers of eigenmodes per configuration. Parameters and nomenclature adopted from Refs.~\cite{ExtendedTwistedMass:2019omo} and \cite{Kotov:2021rah}. Due to excessive computational costs, the eigenmodes for the ensemble with $N_\mathrm{t}=24$ suffer from low statistics. The errors are denoted in the order (scale)(statistical)(systematic), where the total error of the temperatures is given by the scale error only.}
	\label{tab:overview_eigenmodes}
\end{table}

\section{Results}

\begin{figure}
    \centering
	\includegraphics[width=\textwidth]{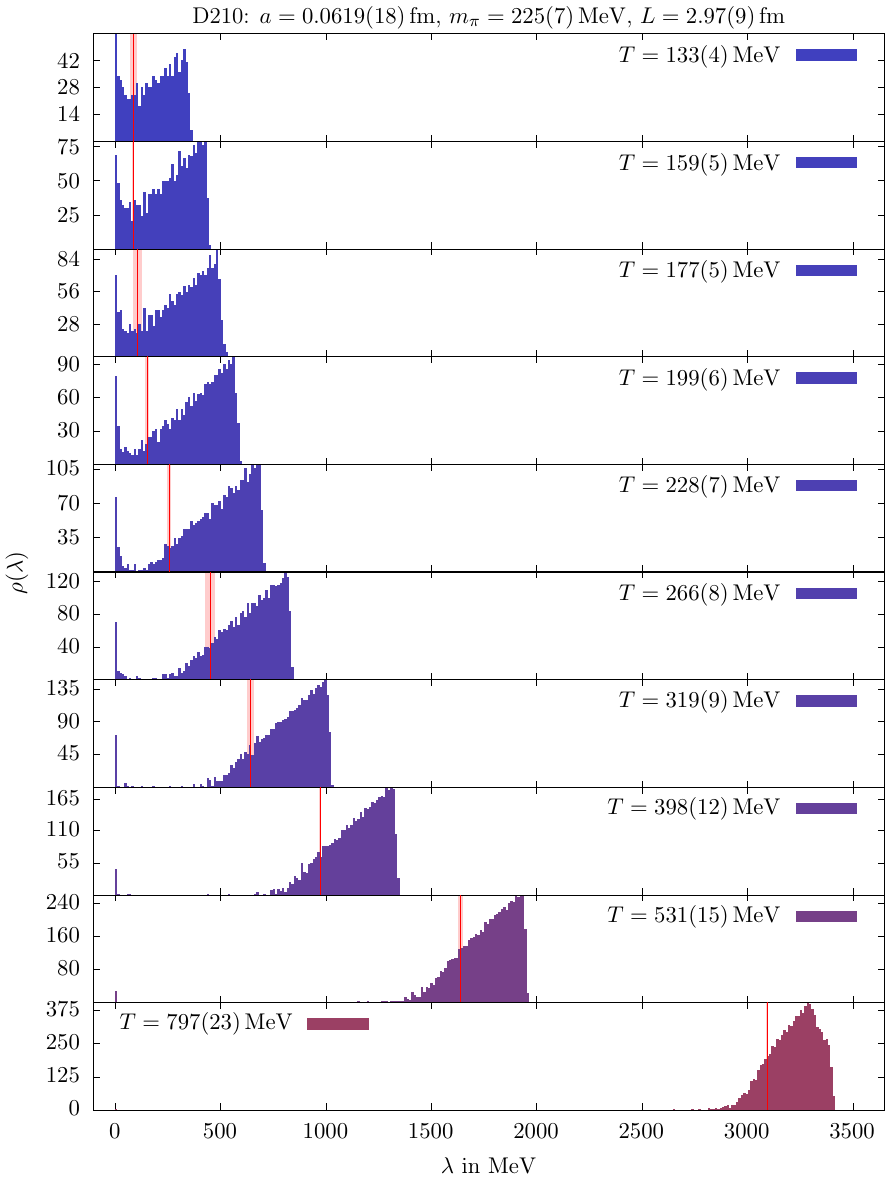}
    \caption{Distributions of (stereographically projected) overlap eigenvalues for different temperatures. The inflection points of the relative volumes (see Figure~\ref{fig:relvol}) are highlighted by vertical red lines, where the shaded regions visualize the statistical together with the systematic error.}
    \label{fig:dist}
\end{figure}

\begin{figure}
    \centering
    \includegraphics[width=\textwidth]{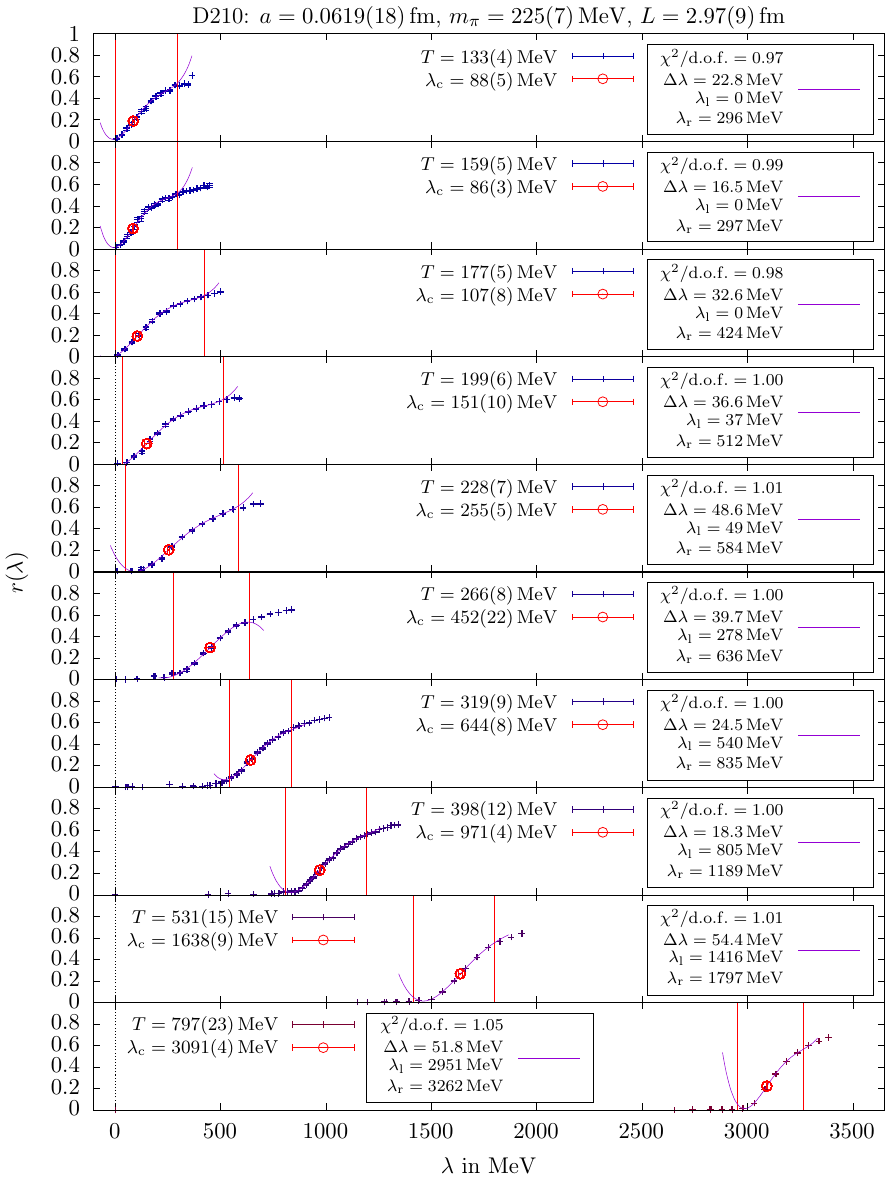}
    \caption{Bin-averaged relative eigenmode volumes as measure of localization for different temperatures. The inflection points are highlighted by red circles, where only the corresponding statistical error is shown. The fit window $[\lambda_{\mathrm l}, \lambda_{\mathrm r}]$ is indicated by vertical red lines and noted together with the binsize $\Delta\lambda$ in the boxes.}
    \label{fig:relvol}
\end{figure}

Figures~\ref{fig:dist} and \ref{fig:relvol} show the distributions of the (stereographically projected) overlap eigenvalues and the bin-averaged relative eigenmode volumes for for each temperature. The mobility edge proxy $\lambda_\mathrm{c}$ is highlighted with a vertical red line in the distribution respectively a red circle in the relative volume plots. As expected the near-zero density of eigenmodes vanishes in high temperature phase according to the Banks-Casher relation
\begin{equation}
	\langle \overline\psi \psi \rangle = -\uppi \, \lim_{\lambda\to0} \lim_{m\to0} \lim_{V\to\infty} \rho(\lambda) \,,
	\label{eq:banks_casher_relation}
\end{equation}
which connects the quark condensate to the spectral density near zero.
Due to the crossover nature of the chiral transition for non-vanishing the near-zero modes do not instantly disappear right above $T_\mathrm{pc}$ but rapidly decrease. 

\begin{figure}
	\centering
	\includegraphics[width=\textwidth]{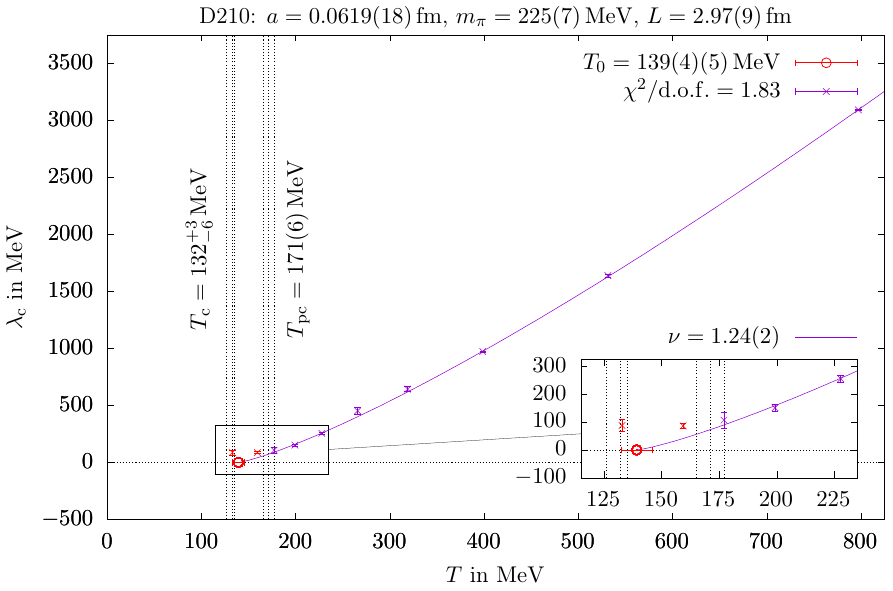}
	\caption{Temperature dependence of the mobility edge as extracted from the bin-averaged relative eigenmode volume. The zero $T_0$, highlighted with the red circle, is estimated by extrapolations with \eqref{eq:crit_scale}. Since the scale error is one-to-one correlated for all data points, just the statistical and systematic error is included for fitting. The scale error of the final result $T_0$ is however denoted in the first bracket and the total error is visualized by the red bar. The critical exponent $\nu$ comes along with a statistical error only. The data points in red were excluded from fit due to the strong deviation.} 
	\label{fig:mobility_edge}
\end{figure}

In Ref.~\cite{Kehr:2023wrs}, the vanishing of the mobility edge is predicted to coincide with the chiral phase transition temperature $T_\mathrm{c}=132^{+3}_{-6}\,\mathrm{MeV}$ (from \cite{HotQCD:2019xnw} and confirmed by \cite{Kotov:2021rah}) in the chiral limit. In contrast to the quadratic extrapolation including the new data points with $N_\mathrm{t} \in \{14,16,18\}$ favors a scaling fit of the form   
\begin{equation}
	\lambda_\mathrm{c}(T) = b (T-T_0)^\nu \,, \label{eq:crit_scale}
\end{equation}
where $T_0$ denotes the Anderson transition temperature and $\nu$ some critical exponent. The temperature dependence of the $\lambda_\mathrm{c}$ together with the extrapolation is shown in Figure~\ref{fig:mobility_edge}. The extrapolated zero at $T_0 = 139(4)(5)\,\mathrm{MeV}$ agrees within errors with the preceding extrapolation of \cite{Kehr:2023wrs} but is slightly higher, which is for the most part explained by the slightly decreased updated lattice spacing. The most recent data points with $N_\mathrm{t} \in \{20,24\}$ marked in red were however excluded from the fit due to the strong deviation. Contrary to the fit prediction the mobility edge does not further decrease and eventually vanish here. This seems like a contradiction to Ref.~\cite{Giordano:2022ghy}, where it was argued that in the chiral limit no goldstone bosons exist if near-zero modes are localized. Assuming that $T_0$ increases, as it happens for $T_\mathrm{pc}$ compared to $T_\mathrm{c}$, when chiral symmetry is broken explicitly, the lower bound of localization should therefore be given by $T_\mathrm{c}$.  

This brings into play a scenario that has already been speculated about in \cite{Kehr:2023wrs}. Even though the relative eigenmode volume as defined in Eq.~\eqref{eq:eigenmode_relvol} is low, it might scale according to some effective dimension when increasing the volume. In this case, the eigenmode would actually not be localized but our choice of localization measure is not sensitive to that. Instead, the volume dependence has to be studied, which was done for the case of quenched QCD in Ref.~\cite{Alexandru:2021pap} and proceeding \cite{Alexandru:2023xho}. Indeed, a second mobility at $\lambda_\mathrm{IR}=0$ was observed for temperatures above $T_\mathrm{IR} \in (200, 250)\,\mathrm{MeV}$ with the zero modes being delocalized. If one assumes that $\lambda_\mathrm{IR}=0$, in contrast to the decreasing $\lambda_\mathrm{c}$, rises when reducing the temperature, both mobility edges would hit and annihilate each other. This speculation gets supported by the observation that $\lambda_\mathrm{c}$ remains constant within errors for temperatures of approximately $T_\mathrm{pc}$ and below. This is also roughly the temperature region where $\lambda_\mathrm{c}$ enters the infrared part of the spectrum as visualized in Fig.~\ref{fig:dist}. As a consequence, $r(\lambda)$ might be inappropriate as measure of localization for these temperatures and also the extrapolation seems to be invalid when chiral symmetry is explicitly broken by non-zero quark masses. However, it is still plausible that the critical scaling according to Eq.~\eqref{eq:crit_scale} applies in a world without near-zero modes above the chiral transition like in the chiral limit with the mobility edge vanishing at $T_\mathrm{c}$. Finally, also possible lattice artifacts deserve further investigation, since the pion mass is still unphysically large and the volume with $m_\pi L \approx 3.39$ rather small compared to that.

The critical exponent $\nu = 1.24(2)$ significantly differs from the expected one of the three-dimensional unitary Anderson model of $\nu \approx 1.44$ \cite{Giordano:2021qav, Ujfalusi_2015}. There are several reasons why this could be the case. First of all and most importantly, if scaling applies, one cannot expect that the scaling window being that large. For instance, the scaling window of the chiral transition was found to be approximately equal to the range $120\,\mathrm{MeV} < T < 300\,\mathrm{MeV}$ \cite{Kotov:2021rah}. Shrinking the fit window tends to result in an increased $\nu$ but slightly worse $\chi^2 / \mathrm{d.o.f.}$. Alternatively, one can try to incorporate next-to-scaling corrections. However, these have to be modeled, since their analytical form is not known. A naive approach, which results in a linear behavior for high temperatures as observed on coarser lattices, leads to a larger $\nu$ but slightly worse $\chi^2 / \mathrm{d.o.f.}$ as well. Therefore, both approaches indicate that the critical exponent is probably larger but are not the most reliable ones for the current data situation. Furthermore, the data points for high temperatures have to be treated more carefully due to the lower number of lattice sites $N_\mathrm{t}$ in the temporal direction. Finally, as explained above, it could be the case that true scaling only applies in the chiral limit and that non-vanishing quark masses give rise to deviations from that.

\section{Outlook}

In order to probe the existence of an infrared mobility edge, it is necessary to employ other definitions of localization. For instance, the ratio $P_2^{-1}(\lambda) / (P_3^{-1}(\lambda))^{1/2}$ shows a dip at the mobility edge in the $\mathrm{SU}(2)$ Higgs model at finite temperature \cite{Baranka:2023ani}. Indeed, first investigations of this quantity in our QCD setup show intimations of a double dip but however better statistic is required. As discussed, especially the evaluation of different spatial volumes would be desirable in order to determine the effective dimension of the eigenmodes. This would enable a more precise determination of the mobility edge with our current definition of localization, namely by employing a finite-size scaling analysis instead of using the inflection point as a proxy. However, this is currently not feasible due to excessive computational costs.

Also the evaluation of ensembles with lower pion masses and at the same time larger $m_\uppi L$. For this purpose, it is planned to evaluate the twisted mass ensembles from Ref.~\cite{Kotov:2020hzm} with a physical pion mass and $m_\uppi L \approx 3.62$ \cite{Alexandrou:2018sjm}, which are based on the vacuum ensembles from Ref.~\cite{Alexandrou:2018egz}. Due to the number of lattice sites, this comes along with high computational costs as well. 

Therefore, also an accelaration of the computations is required. Very promising for this purpose is the prior smoothing of the gauge configurations using gradient flow as proposed in Ref.~\cite{Luscher:2010iy}, which reduces ultraviolet fluctuations on the lattice. Doing so, one can perform the analysis for different flow times and extrapolate the mobility edge to vanishing flow time afterwards. Furthermore, optimizations on the numerical side like multigrid methods and polynomial preconditioning in order implement the overlap operator in a more efficient way might be worthwile \cite{Brannick:2014vda, Frommer:2024cgc}.

Finally, it would be interesting to study the QCD Anderson transition in an external magnetic field. It is known that in this case the chiral transition as well as the deconfinement temperature decrease with increasing field strength \cite{Bali:2011qj, Bruckmann:2013oba}. If the conjecture that the QCD Anderson transition is connected to both, one would expect that its transition temperature decreases as well, hence, the onset of eigenmode localization gets shifted to lower temperatures. Therefore, studying magnetic fields would be a good way to qualitatively check that conjecture.

\end{document}